\documentclass[11pt]{article}
\pdfoutput=1

\usepackage[T1]{fontenc}
\usepackage[utf8]{inputenc}
\usepackage[english]{babel}
\usepackage[version=4]{mhchem}
\usepackage{subcaption}
\usepackage{multirow}
\usepackage{bm}
\usepackage{lineno}
\usepackage{accents}                      	
\usepackage{ulem}                            	
\usepackage[noblocks]{authblk}		%
\usepackage{tabularx}                        	%
\usepackage{graphicx}                        	%
\usepackage{bm}                        	%
\usepackage{amsmath}                        	%
\usepackage{amssymb}                        	%
\usepackage{float} 				
\usepackage{units}				%
\usepackage{color}				%
\usepackage[dvipsnames]{xcolor}		%
\usepackage{hyperref}			%
\usepackage{url}			%
\usepackage[percent]{overpic}
\usepackage{siunitx}

\newcommand{\diff}{ {\mathrm{d}} }
\newcommand{\feni}{\text{FeNi~}}
\newcommand{\kj}{\text{FeNiSi$_2$O$_4$~}}
\newcommand{\badro}{\text{FeNiSi$_7$O$_2$~}}
\newcommand{\tagawa}{\text{FeNiSiH~}}
\newcommand{\sakamaki}{\text{FeNiH~}}
\newcommand{\flux}{\Phi}

\makeatletter
\def\@maketitle{%
  \newpage
  \null
  \vskip 0.em%
  \begin{center}%
  \let \footnote \thanks
    {\huge \bfseries \@title \par}%
    \vskip .0em%
    {\normalsize
      \lineskip 5.em%
      \begin{tabular}[t]{c}%
        \@author
      \end{tabular}\par}%
    {\normalsize \@date} 
  \end{center}%
  \par
  \vskip .em}
\makeatother

\begin{document}
\normalem 

\author[1]{Lukas~Maderer}
\author[2]{Edouard~Kaminski\thanks{\scriptsize Contact:~\texttt{kaminski@ipgp.fr}}}
\author[1,3]{Jo\~ao~A.~B.~Coelho}
\author[1]{Simon~Bourret}
\author[1,4]{V\'eronique~Van~Elewyck\thanks{\scriptsize Contact:~\texttt{elewyck@apc.in2p3.fr}}}

\affil[1]{\small Universit\'e Paris Cit\'e, CNRS, Astroparticule et Cosmologie, F-75013 Paris, France}
\affil[2]{\small Universit\'e Paris Cit\'e, Institut de Physique du Globe de Paris, CNRS, F-75005 Paris, France}
\affil[3]{\small Universit\'e Paris-Saclay, CNRS/IN2P3, IJCLab, F-91405 Orsay, France}
\affil[4]{\small Institut Universitaire de France, F-75005 Paris, France}

\title{Unveiling the outer core composition with neutrino oscillation tomography}
\maketitle

\abstract{
 In the last 70 years, geophysics has established that the Earth’s outer core is an FeNi alloy containing a few percent of light elements, whose nature and amount remain controversial today. Besides the classical combinations of silicon and oxygen,  hydrogen has been advocated as the only light element that could account alone for both the density and velocity profiles of the outer core. Here we show how this question can be addressed from an independent viewpoint, by exploiting the tomographic information provided by atmospheric neutrinos, weakly-interacting particles produced in the atmosphere and constantly traversing the Earth. We evaluate the potential of the upcoming generation of atmospheric neutrino detectors for such a measurement, showing that they could efficiently detect the presence of 1 wt\% hydrogen in an FeNi core 
in 50 years of concomitant data taking. 
We then identify the main requirements for a next-generation detector to perform this measurement in a few years timescale, with the further capability to efficiently discriminate between FeNiH and FeNiSi$_x$O$_y$ models in less than 15 years.
}

\newpage

\section{Introduction}

The determination of the composition of the Earth core is a long-standing debate in Earth sciences (e.g. ~\cite{birch52,poirier84,hirose2013,mcdonough}). Seismology combined with experimental petrology shows that the presence of light elements is required to account for the density and seismic velocity profiles in the core, e.g., PREM \cite{prem}. But the precise nature of the light elements remains elusive, and various combinations of, e.g., Si and O~\cite{badro,KJ}, can equally fit PREM. Hydrogen, the most abundant element in the proto-solar nebula, is the only candidate that could account both for the density and velocity profiles in the core without the need for any additional light element~\cite{umemoto2015, sakamaki2016,yuan}. An FeNiH core, rather than a more classical FeNiSi$_x$O$_y$ composition, would bear profound consequences for Earth formation scenarios and modeling of convection and generation of the magnetic field in the outer core. The aim of the present study is to show how the development of high-performance detectors of atmospheric neutrinos opens a new path to independently constrain the composition of the core and test the FeNiH hypothesis in particular.

Neutrinos are neutral elementary particles that exist in three flavors: electron~($e$), muon~($\mu$) and tau~($\tau$). Because they feel only the weak interaction, they can traverse and emerge from very dense media, including the Earth itself. The interactions of cosmic rays with the atmosphere generate an abundant and ubiquitous flux of energetic neutrinos, which undergo flavor oscillations while crossing the Earth~\cite{SKosc}. The probability of their flavor transition depends on the neutrino energy and path length, but also on the density of electrons $n_e$ in the traversed medium~\cite{MikheyevSmirnov, Wolfenstein}. Because of its dependence in $n_e$, this effect is key in accessing chemical properties of the traversed medium through neutrino oscillation physics. Assuming an Earth's radial structure and composition, $n_e$ as a function of $r$, the radial distance from the center of the Earth, is given by 
\begin{equation}
\label{eq:nue}
n_e(r) = \frac{Z}{A}(r) \times \rho(r) \times \mathcal{N}_A 
\end{equation}
where $\mathcal{N}_A$ is the Avogadro number, $\rho$ is the mass density, and $Z/A$ is the proton-to-nucleon ratio defined as:
\begin{equation}
\label{eq:Z/A}
\frac{Z}{A} = \sum_i w_i \frac{Z_i}{A_i}
\end{equation} 
where $Z_i$, $A_i$ and $w_i$ are respectively the local  atomic number, standard atomic weight and weight fraction of element $i$ of the material. 
Chemical elements with different Z/A will therefore generate distinct signals in the neutrino oscillation pattern for a given mass density. The signature of these oscillations in the atmospheric neutrino signal observed at a detector can thus reveal the nature of the matter they interacted with.

While this method of Earth tomography has been conceptually explored in different contexts since the 1980s (see e.g. ~\cite{Ermilova86,Nicolaidis88,Tarantola91,Ohlsson:2001ck, Ohlsson:2001fy,Lindner:2002wm,winter2006} and references therein), only now does the upcoming generation of atmospheric neutrino detectors provide a concrete opportunity to evaluate its capability of probing the density and/or composition of the deep Earth. Oscillation tomography with atmospheric neutrinos has been recently discussed by several authors~\cite{rott,winter,bourretICRC,bourretVLVnT,DOlivo,INOtomography,DUNEtomography}, but its relevance and potential for the characterization of the Earth's core composition has not been systematically explored and discussed yet from the Earth sciences point of view. We propose to do so in this study, with a specific focus on the capacity of this method to tackle the question of Hydrogen in the core, thanks to the very large Z/A of H compared to Fe and other light elements. 

Section~\ref{sec:results} presents our main results on neutrino tomography of the Earth's outer core. In section~\ref{subsec:res_theo} we start by quantifying the expected theoretical reach of the method under the hypothesis of perfect detector capabilities, showing how different assumptions for the outer core composition will indeed modify the number of neutrinos of different flavours interacting at a detector site. In section~\ref{subsec:res_currgen} we move to investigating the realistic performances of the two main families of atmospheric neutrino detectors presently under construction, and discuss their ability to detect the presence of 1wt\% Hydrogen in the outer core. Finally, we illustrate in section~\ref{subsec:res_nextgen} the evolution required for the next generation of detectors to be able to discriminate FeNiH vs. FeNiSi$_x$O$_y$ models of the Earth's core. The methods and inputs used for the computations that support the present study are described in detail in section~\ref{sec:methods}. The context and implications of our findings are further discussed in section~\ref{sec:discussion}.


\section{Neutrino tomography of the Earth's outer core}
\label{sec:results}

\subsection{Theoretical approach}

\label{subsec:res_theo}

Our first step is to evaluate the intrinsic sensitivity of the method by conducting detailed computations of the propagation of neutrinos through the Earth and comparing the expected rate of events in a perfect detector under different assumptions for the outer core composition. In this section we limit ourselves to the key theoretical concepts of the method, required to understand its application to the tomography of the core; a more detailed presentation of neutrinos oscillations and propagation in matter is provided in section~\ref{sec:methods}.

\begin{figure}[hb!]
\includegraphics[width=.45\linewidth]{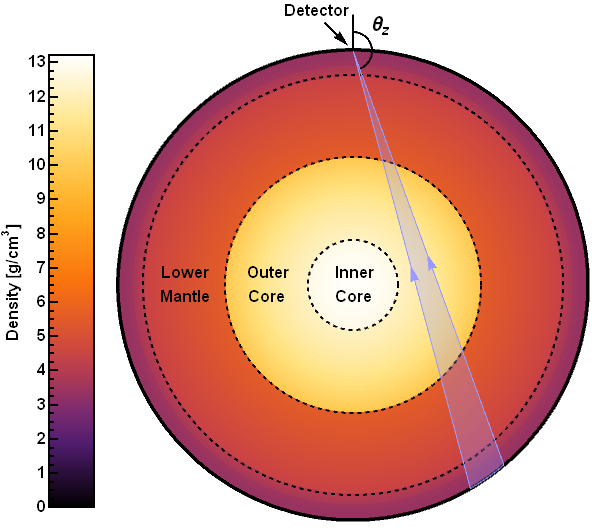}
\includegraphics[width=.52\linewidth]{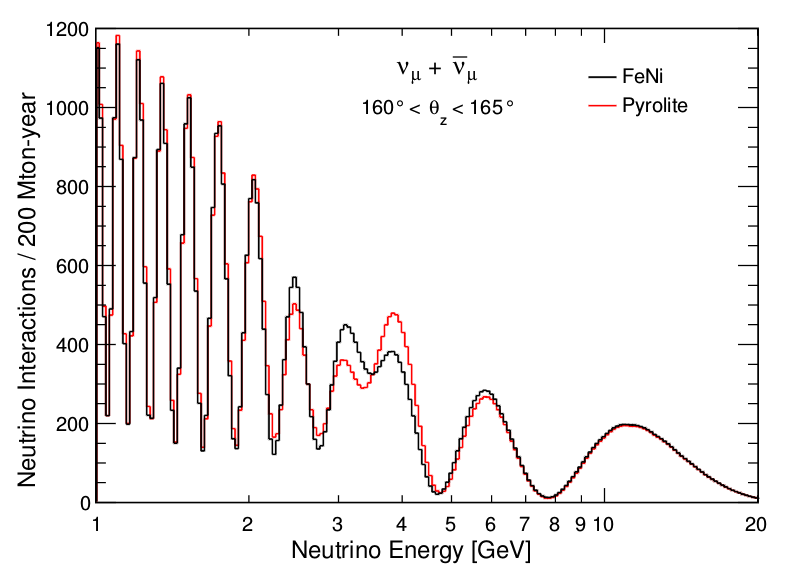}
\caption{\textbf{Oscillations of Earth-crossing neutrinos. Left:} A detector located at the surface will receive atmospheric neutrinos having traversed the Earth along different paths, as measured by their zenith angle $\theta_z$ with respect to the vertical at the detector location. As these neutrinos cross the Earth, the amplitude of their oscillations at different depths will be affected by the local density of electrons. A modification in the electron density of the outer core would then affect the flavor mix of neutrinos that cross it. \textbf{Right:} To illustrate this effect, the number of interactions expected for atmospheric muon neutrinos and antineutrinos ($\nu_\mu+\bar{\nu}_{\mu}$) is shown as a function of the neutrino energy in a detector with 200 Mton-years exposure, assuming two extreme chemical compositions chosen for the sake of the argument, corresponding respectively to pure FeNi (a light-element-free outer core) and to pyrolite (same as the mantle).  Only neutrinos expected from the angular region  $\theta_z \in [160^\circ, 165^\circ ]$ highlighted on Panel (a), with a large intersection with the outer core, are shown. The observed oscillation pattern reflects a combination of disappearing ($\nu_\mu\rightarrow\nu_e, \nu_\tau$) and appearing ($\nu_e \rightarrow\nu_\mu$) muon-neutrinos. Differences related to the outer core composition model are most visible for neutrino energies between 2 and 5~GeV.}
\label{fig:OscExample}
\end{figure}

When they cross the Earth, atmospheric neutrinos experience resonant oscillations at a depth that depends on their energy and on the electron density $n_e(r)$ as defined in equation~\ref{eq:nue}. For core-crossing neutrinos, this resonance happens at energies around 3 GeV, as further discussed in section~\ref{subsec:nuosc}. For a given flavor $\alpha\,(=e,\mu,\tau)$ observed at the detector, the information on $n_e(r)$ is encoded in the rate of interacting events $R^{int}_\alpha (E,\theta_z)$:

\begin{equation}
\label{eq:rate}
    R^{\text{int}}_{\alpha}(E, \theta_z) \equiv 
    \frac{\diff^4 N^{\text{int}}_{\alpha}(E, \theta_z)}{\diff E\, \diff\theta_z\,\diff t\, \diff M} = 
    \left(\sum_{\beta=e,\mu} \frac{\diff^2 \flux_{\nu_\beta} }{\diff E\, \diff\theta_z} (E,\theta_z) \cdot P_{\nu_\beta \to \nu_\alpha}(E, \theta_z) 
  \right) \cdot \frac{\sigma^{\text{int}}_{\nu_\alpha}(E)}{m_N} 
\end{equation}
 where we have made explicit the dependence on the neutrino energy $E$ and on incoming zenith angle $\theta_z$, as defined in figure~\ref{fig:OscExample} (left), under the assumption of a spherically symmetric mass density profile of the Earth. The angle $\theta_z$ is directly related to the path length $L$ through the Earth: $L\approx-2 R_\oplus \cos{\theta_z}$ with $R_\oplus$ the radius of the Earth. Extracting tomographic information on $n_e(r)$ in a given Earth layer therefore requires a detailed knowledge of the flavor, path length and energy distribution of neutrinos having crossed that layer. 

In this equation, $R^{\text{int}}_{\alpha}(E, \theta_z)$ represents the differential rate of $\nu_\alpha$ interactions (with target nucleons of mass $m_N$) at the detector location, as a function the energy and zenith angle and per unit exposure (defined as the product of running time and target mass of the detector). It is obtained as a product of the incident (differential) fluxes of atmospheric neutrinos $\Phi_\beta$ (with $\beta=e,\mu$ as the atmospheric $\nu_\tau$ component is negligible), the flavor oscillation probability along each neutrino path $P_{\nu_\beta \to \nu_\alpha}$, and the neutrino-nucleon cross-section $\sigma^{int}_{\nu_\alpha}$, that quantifies the probability for a neutrino to interact (hence to generate a potentially detectable signal). A quantitative example is shown in the right panel of figure~\ref{fig:OscExample}, illustrating  both the oscillatory effects and the attenuation of the flux at high energies, which is due to the energy power spectrum $\propto E^{-3}$ of the cosmic ray flux that produce the neutrinos in the atmosphere~\cite{Honda}. 

From equation (\ref{eq:rate}), one can generate the two-dimensional (2D)  distributions of expected number of neutrino interactions of a given flavor as a function of $(E, \theta_z)$, or {\it oscillograms}, for any specific exposure of a perfect detector that would observe atmospheric neutrinos from all directions. Because of the dependence of neutrino oscillation on $n_e$ along the neutrino path, changes in the outer core composition (i.e., different Z/A) induce variations in the expected signal, depending on the neutrino energy and exact trajectory (or equivalently, $\theta_z$), that get imprinted in the oscillograms. 

\begin{figure}[ht!]
\centering
\includegraphics[width=.49\linewidth]{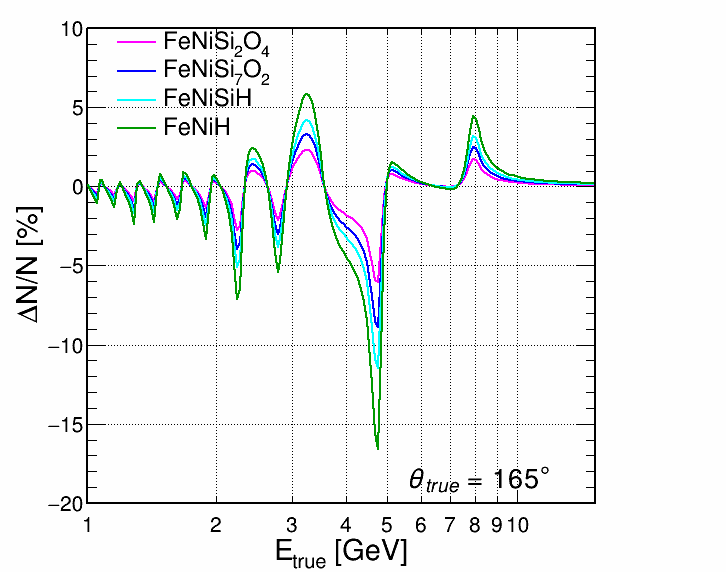}
\includegraphics[width=.49\linewidth]{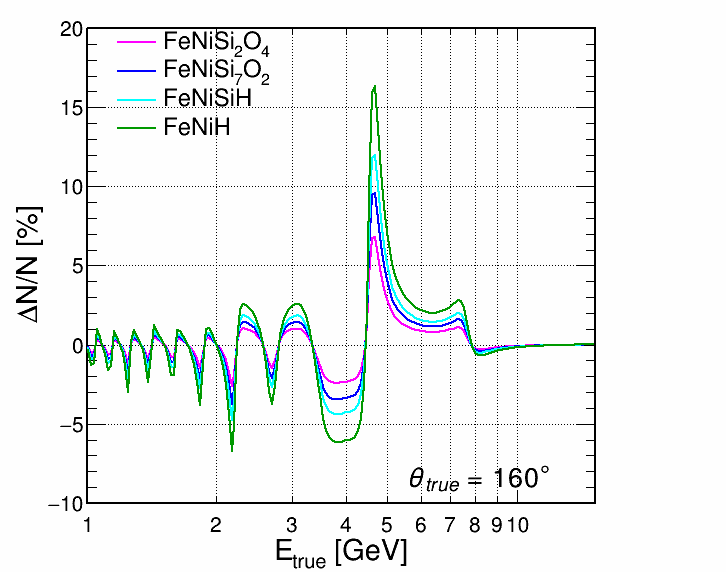}
\\
\includegraphics[width=.49\linewidth]{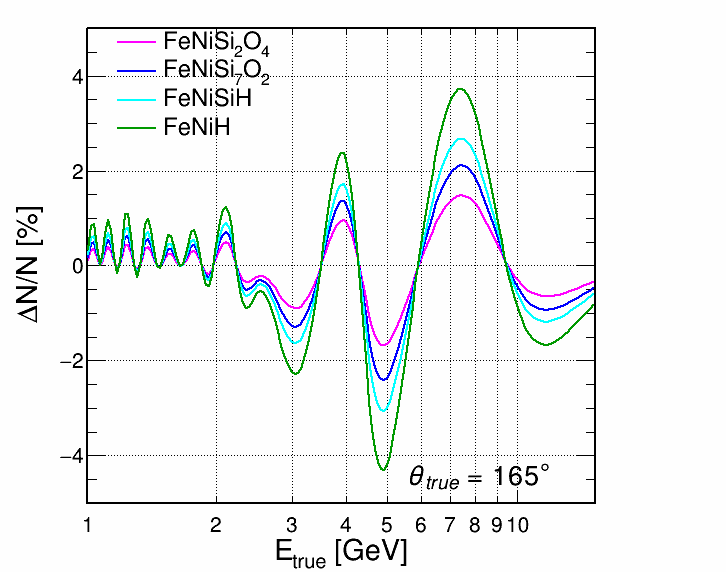}
\includegraphics[width=.49\linewidth]{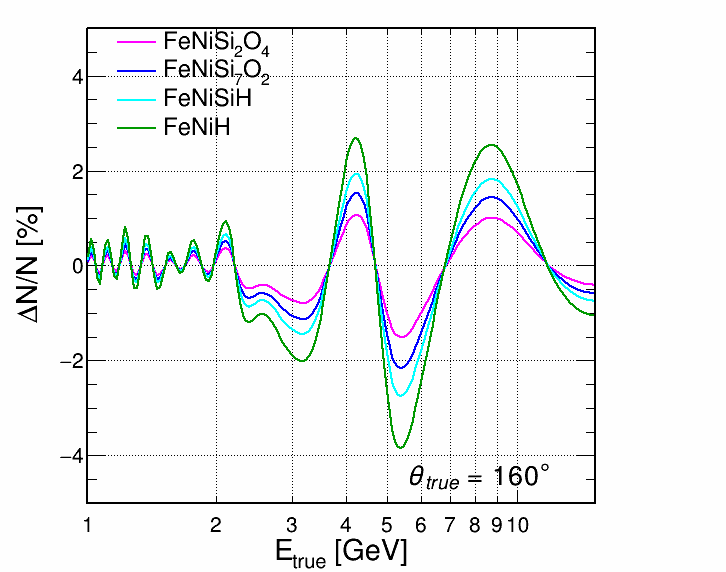}
\caption{ {\bf Predicted relative difference $\Delta\! N/N$ in the number of neutrino interactions as a function of the incoming (or ``true") neutrino energy, for the different outer core compositions described in Tab.~\ref{tab:Z/A}, taking \feni as a reference.} Upper panels  are for $\nu_\mu+\bar{\nu}_{\mu}$, while lower panels are for $\nu_e+\bar{\nu}_{e}$. Two incoming neutrino directions have been considered:  $\theta_z = 165^\circ$ (left) and $ 160^\circ$ (right).}
\label{fig:1D_DIFF_PERFECT_allmodels}
\end{figure}

We have quantified this effect by computing the relative difference ($\Delta\! N/N$) in expected number of neutrino events between oscillograms generated with different models of outer core composition. We considered five different models of outer core composition (see table~\ref{tab:Z/A}), ranging from a pure \feni alloy to a Hydrogen-rich (\sakamaki\!) one. 
figure~\ref{fig:1D_DIFF_PERFECT_allmodels} illustrates the expected impact of the composition in terms of $\Delta\! N/N$ for muon- and electron-neutrino interactions, as a function of the neutrino energy, for two different incoming directions. 
The predicted numbers of interactions for those particular configurations are shown to differ by up to 4\% for electron neutrinos, and up to 20\% for muon neutrinos, depending on the core composition considered. Furthermore, because the shape of the signal changes a lot depending on the zenith angle, it appears important to consider the signal in two dimensions (i.e. as a function of $E$ and $\theta_z$). This is further illustrated in figure~\ref{fig:2D_DIFF_FeNIvsFeNiH_perfect_NextGen} showing the full 2D distributions of $\Delta\! N/N$ for muon- and electron-neutrino interactions, for a specific comparison between FeNiH and FeNi core compositions. Such results confirm that a detailed study of the neutrino event rate in the energy range 1 - 10 GeV has the potential to constrain the Earth's outer core composition.

\subsection{Sensitivity of upcoming neutrino detectors to the outer core composition}
\label{subsec:res_currgen}

\begin{figure}[ht!]
\centering
\includegraphics[width=.49\linewidth]{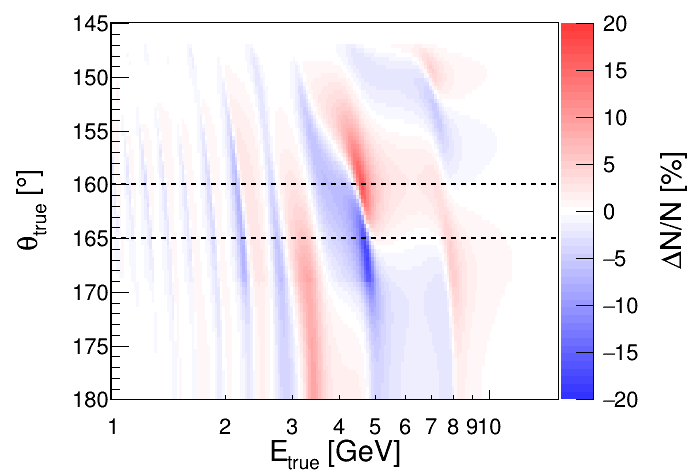}
\includegraphics[width=.49\linewidth]{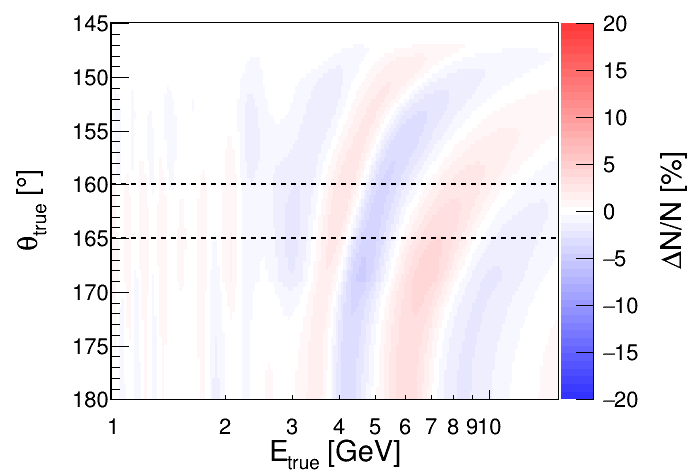}
\\
\includegraphics[width=.49\linewidth]{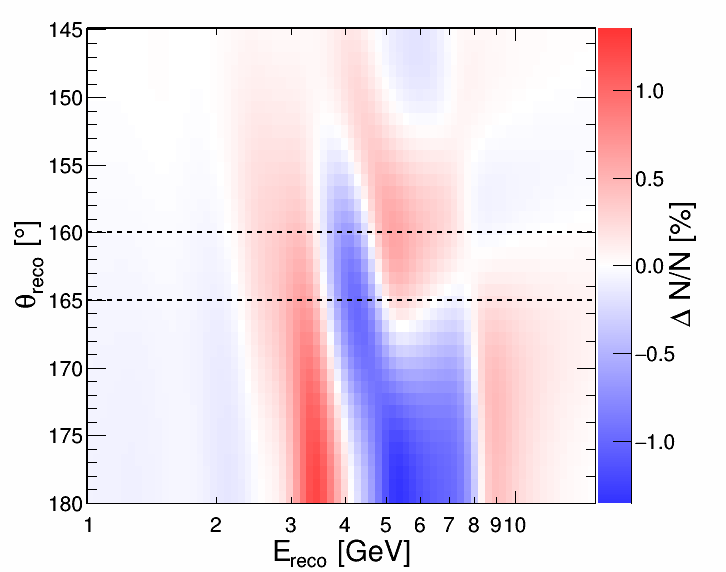}
\includegraphics[width=.49\linewidth]{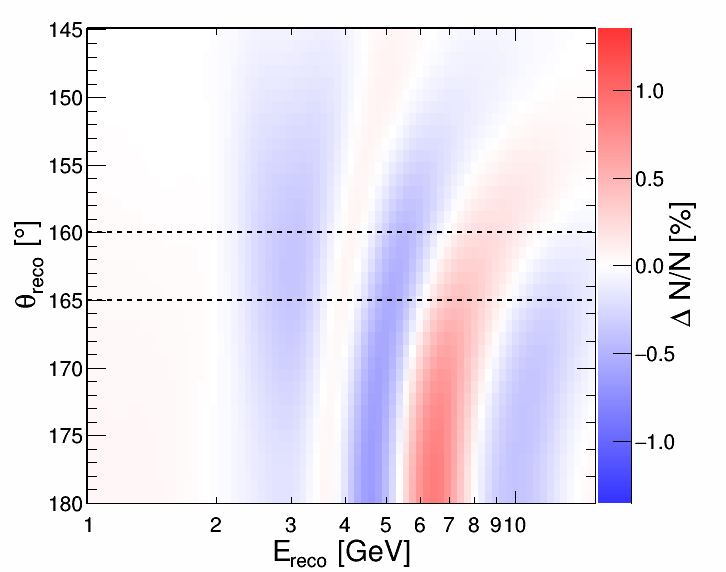}
\caption{\textbf{Upper panels: expected relative difference $\Delta\! N/N$ in the number of  neutrino interactions,  as a function of the incoming (or ``true") neutrino energy and zenith angle, for the composition models \sakamaki vs. \feni.} Left panel is for $\nu_\mu+\bar{\nu}_{\mu}$, while right panel is for $\nu_e+\bar{\nu}_{e}$. The horizontal dashed lines correspond to the two incoming directions considered for the plots in figure~\ref{fig:1D_DIFF_PERFECT_allmodels}. {\bf Lower panels: expected relative difference $\Delta\! N/N$ in the number of detected events, as a function of the reconstructed neutrino energy and zenith angle, for the composition models \sakamaki vs. \feni, for the Next-Generation detector described in section~\ref{subsec:detectors}  and table~\ref{tab:detectors}).} Left panel is for the {\it track}-like events (mostly associated to $\nu_\mu$  interactions) and right panel is for {\it cascade-}like events (mainly from $\nu_e$ and $\nu_{\tau}$ interactions), as defined in section~\ref{subsec:detectors}. 
}
 \label{fig:2D_DIFF_FeNIvsFeNiH_perfect_NextGen}
 \end{figure}

 The distributions shown in the upper panels of figure~\ref{fig:2D_DIFF_FeNIvsFeNiH_perfect_NextGen} implicitly relate to a detector with 100\% detection efficiency, perfect energy and $\theta_z$ resolution, and infinite statistics. In reality, the measurement accuracy will be limited by experimental effects, resulting in some blurring and attenuation of the signal, as illustrated in figure~~\ref{fig:2D_DIFF_FeNIvsFeNiH_perfect_NextGen} (lower panels). The detector technology and specifications (size, detection efficiency, resolution and particle identification) affect its ability to reconstruct the neutrino properties (energy, direction and flavor). Additionally, the intrinsically probabilistic nature of neutrino interactions as a quantum process induces a statistical uncertainty on the final measurement (in terms of observed number of events). The detector thus needs to be scalable to a sufficiently large volume for this uncertainty to be smaller than the intrinsic signal.  

Neutrinos are observed only indirectly, through the signal deposited in the detector by the byproducts of their interaction. The ability to detect, reconstruct an identify the different types of neutrino events in the energy range of interest for Earth tomography is essentially driven by the  detector size, technology and density of sensors. We consider here two main experimental  approaches that are currently pursued for the upcoming generation of neutrino detectors at the GeV scale, and that rely on different target materials and  observation strategies. Liquid Argon (LAr) detectors, such as DUNE~\cite{DUNE-TDR}, are a type of Time Projection Chamber (TPC)~\cite{Rubbia:1977} that detects the electrons released upon ionization of Argon atoms along the paths of the secondary charged particles emerging from the neutrino interaction. Water-Cherenkov (wC) detectors instrument large volumes of water (or ice) with 2D -- such as Hyperkamiokande~\cite{HyperK1, HyperK2} -- or 3D -- such as ORCA~\cite{KM3NeTLoI,ORCA}-- arrays of photosensors that detect the Cherenkov light induced by the secondary charged particles traveling faster than light in water~\cite{cherenkov,FrankTamm}. As further discussed in section~\ref{subsec:detectors}, LAr-detectors reach the highest reconstruction precision, while wC-detectors are scalable to larger volumes, especially when deployed in a natural environment as is the case of ORCA. 

To assess the impact of these experimental limitations on the sensitivity of neutrino oscillation tomography, we have modeled the detector response with four key features: the effective mass -- defined as the product of the instrumented target mass and the detection efficiency --, the direction and energy resolutions  -- which characterize the mapping of the neutrino true energy and zenith angle $(E,\theta_z)$ into the respective reconstructed quantities $E_\text{reco}$ and  $\theta_\text{reco}$ --, and the efficiency in identifying the neutrino flavor based on the secondary particles produced in the interaction. In this study we assume that all detectors under consideration have the capability to classify events into two main observational classes: {\it track}-like events, which are mostly associated to $\nu_\mu$ charged-current interactions producing a long muon track, and {\it cascade}-like events, when there is no such single muon or the muon energy is too small for it to be distinguished among the many other secondary particles created in the event. The {\it cascade} channel therefore comprises mainly $\nu_e$ and (subdominantly) $\nu_\tau$ interactions, plus a small contribution from $\nu_\mu$ interactions without an observable muon. More detail about the mapping of neutrino interaction types into the observational {\it track} and {\it cascade} classes is provided in section~\ref{subsec:detectors}.  

To investigate the influence of the detector characteristics on their potential to constrain the core composition, we consider four benchmark parametrizations: three with the expected capabilities of the detectors under construction (DUNE, ORCA, and HyperKamiokande), and one additional, hypothetical ``Next-Generation" (NextGen) detector with improved performances. The corresponding modelling, and the physical parameters used for each specific detector, are described in section~\ref{subsec:detectors} and table~\ref{tab:detectors}. By folding the detector response with the expected rate of neutrino interactions given by equation (\ref{eq:rate}), and summing over all relevant interaction types, we obtain the predicted rate of observable events of a given observational class, $R^\text{obs}_\text{track}(E_\text{reco},\theta_\text{reco})$ or $R^\text{obs}_\text{cascade}(E_\text{reco},\theta_\text{reco})$ in a given bin of neutrino reconstructed energy and zenith angle. By integrating these quantities over time and effective mass of the detector, we generate 2D histograms of the expected number of events as a function of $E_\text{reco}$ and $\theta_\text{reco}$, for a given detector exposure. Such oscillograms are the cornerstone of our discussion of the detectors capability to constrain the Earth's core composition, as illustrated in the lower panels of figure~\ref{fig:2D_DIFF_FeNIvsFeNiH_perfect_NextGen}. 

\begin{figure}[]
\centering
\begin{overpic}[width=.49\linewidth]{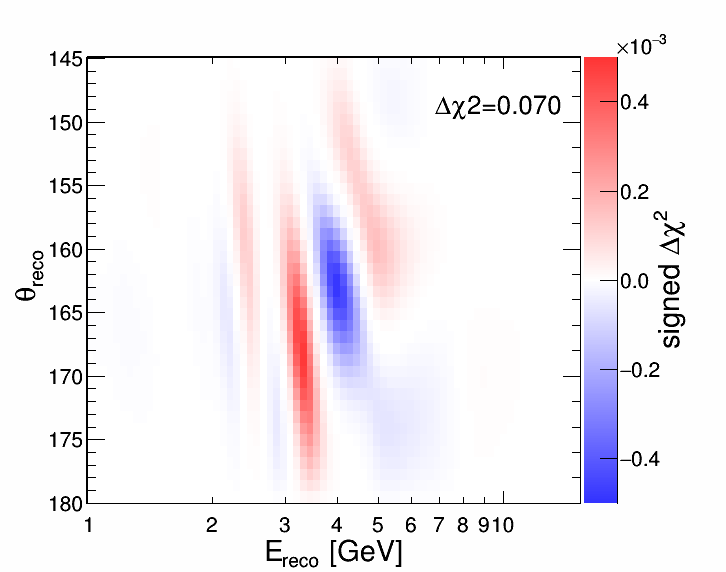}
\put (15,11) {DUNE,~{\it tracks}}
\end{overpic}
\begin{overpic}[width=.49\linewidth]{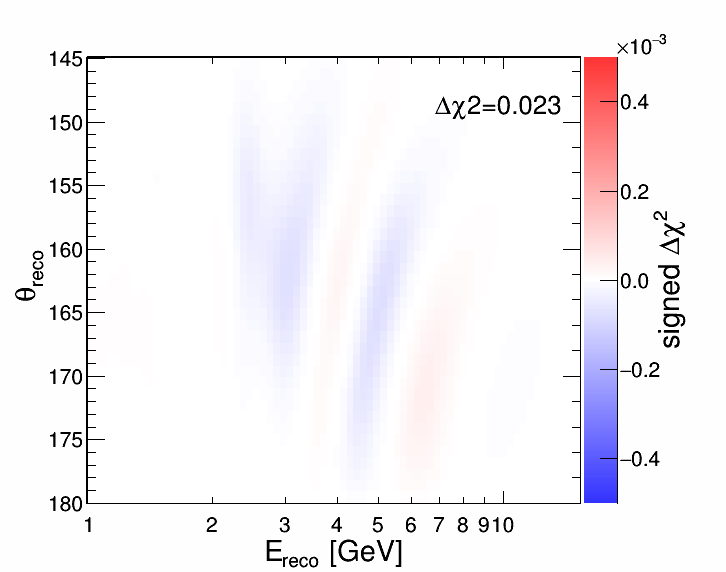}
\put (15,11) {DUNE,~{\it cascades}}
\end{overpic}
\begin{overpic}[width=.49\linewidth]{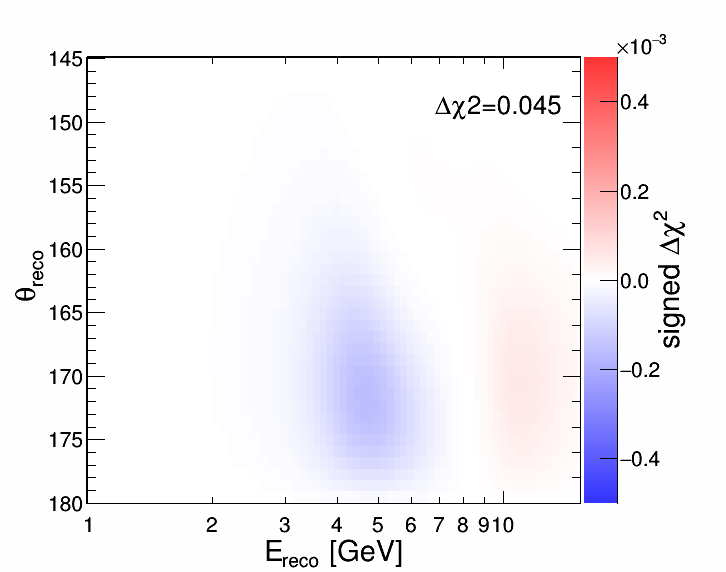}
\put (15,11) {ORCA,~{\it tracks}}
\end{overpic}
\begin{overpic}[width=.49\linewidth]{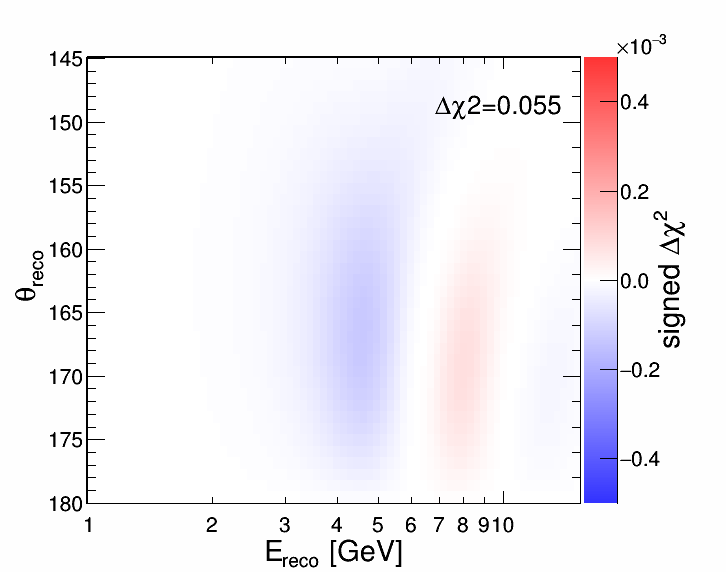}
\put (15,11) {ORCA,~{\it cascades}}
\end{overpic}
\begin{overpic}[width=.49\linewidth]{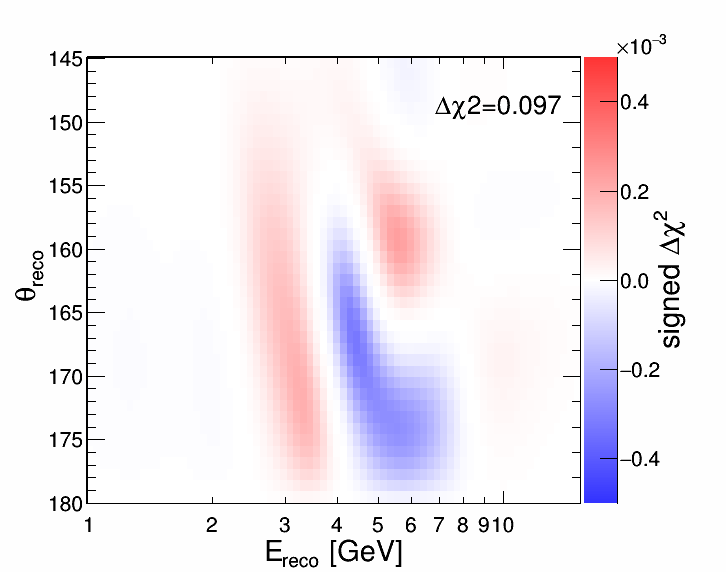}
\put (15,11) {HyperK,~{\it tracks}}
\end{overpic}
\begin{overpic}[width=.49\linewidth]{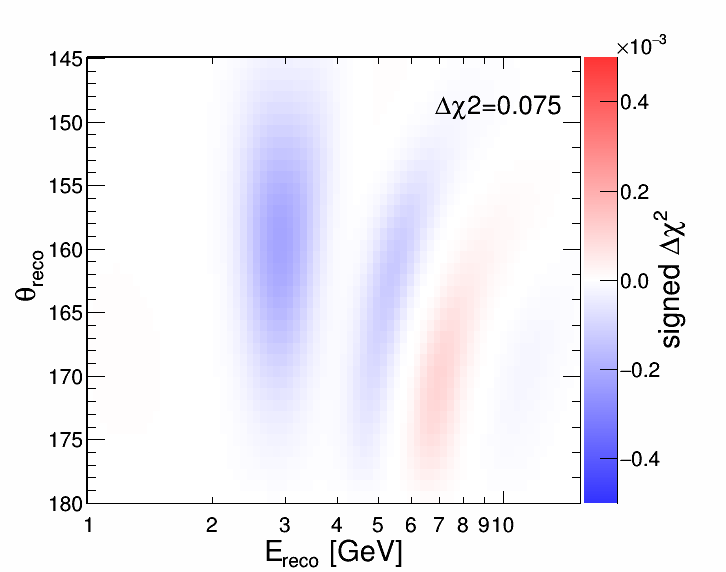}
\put (15,11) {HyperK,~{\it cascades}}
\end{overpic}
\caption{{\bf $\Delta\chi^2$ sensitivity for discriminating between FeNi model and \sakamaki model in 20 years livetime of upcoming detectors.} From top to bottom, the panels show the signed $\Delta \chi^2$ maps (as defined in Section~\ref{subsec:stat}) as a function of the reconstructed neutrino energy and zenith angle, for DUNE, ORCA, and HyperKamiokande, for {\it track}-like events \textbf{(left column)} and {\it cascade}-like events \textbf{(right column)} as defined in section~\ref{subsec:detectors}. The color scale is the same in all plots. The number indicated in each plot corresponds to the total $\Delta\chi^2$ sensitivity summed (in absolute value) over all bins of the $(E_{reco},\theta_{reco})$ plane.} 
\label{fig:CHI2_UPCOMINGDETS_FeNivsFeNiH}
\end{figure}

For a given detector, our projected experimental sensitivities are obtained from the detailed comparison of the full 2D oscillograms of expected neutrino events, generated with different assumptions for the core composition. A statistically meaningful way of quantifying the detector performance, described in section~\ref{subsec:stat}, is to apply a $\chi^{2}$ hypothesis test to the 2D histograms of expected events in bins of $(E_{reco},\theta_{reco})$. The total $\Delta\chi^2$ associated to a given pair of composition models, say A and B, which under the assumption of the validity of Wilk's theorem is directly related to the significance $\sigma = \sqrt{\Delta\chi^2}$, giving the confidence level (C.L.) by which model A can be discriminated from model B. 
Examples of signed $\Delta\chi^2$ maps for different upcoming detectors are presented in figure~\ref{fig:CHI2_UPCOMINGDETS_FeNivsFeNiH} for the discrimination between \feni and \sakamaki models. The three detectors achieve a comparable statistical significance in the measurement of the outer core Z/A after 20 years of data taking, although their sensitivity does not necessarily come from the same region of the $(E,\theta)$ plane, nor from the same observational channel. 

\begin{figure}[ht!]
\centering
\includegraphics[width=.49\linewidth]{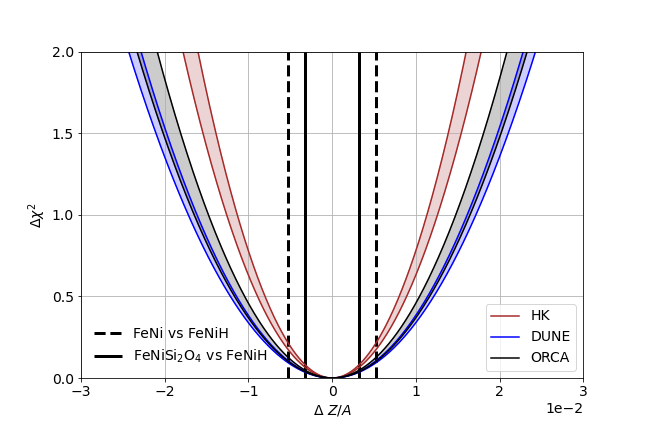}
\includegraphics[width=.49\linewidth]{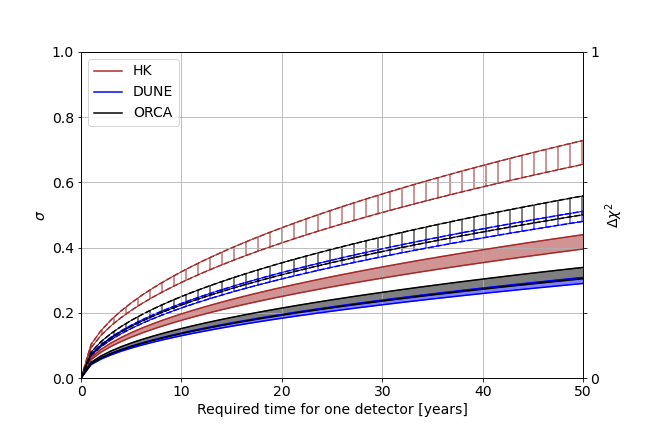}
\caption{\textbf{Sensitivity of the individual upcoming detectors to the outer core composition. Left:} Sensitivity profile for the absolute precision in the Z/A measurement, for 20 years operation of the HyperKamiokande, DUNE and ORCA detectors. The vertical lines indicate the Z/A separation between specific pairs of models of core composition. \textbf{Right:} Sensitivity bands as a function of the detector livetime, for discriminating specific pairs of models. In the right plot, the filled bands represent the discrimination power between \kj and \sakamaki\!, while the hashed bands represent \feni vs \sakamaki\!. The band limits in both plots  correspond to the baseline and  optimistic assumptions on the detectors capability to filter out events which are insensitive to neutrino flavor (as explained in section~\ref{subsec:detectors}).}
\label{fig:CHI2EVOL_UPCOMINGDETS}
\end{figure}

\begin{figure}[ht!]
\centering
\includegraphics[width=.49\linewidth]{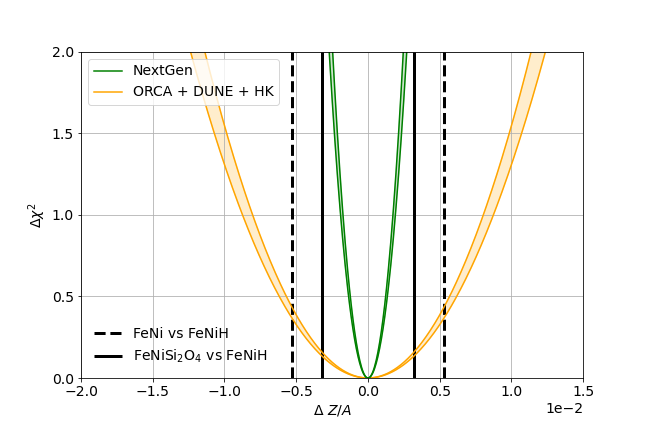}
\includegraphics[width=.49\linewidth]{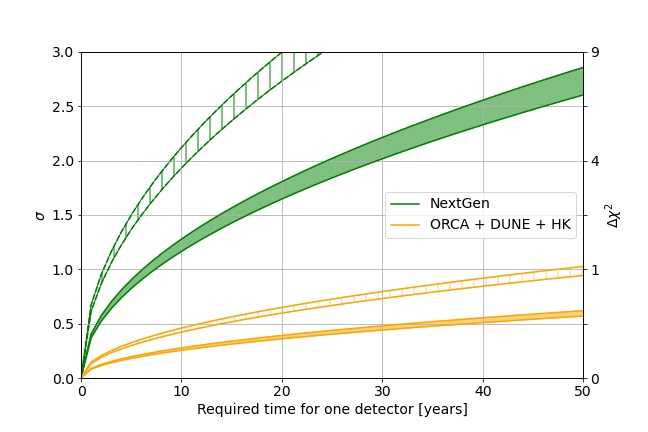}
\caption{{\bf Sensitivity to the outer core composition for upcoming and next-generation detectors.}  \textbf{Left:} $\Delta\chi^2$ profile for the absolute precision in the Z/A measurement, for 20 years operation of a Next Generation (NextGen) detector and for the combination of DUNE, ORCA and HyperKamiokande (HK) over the same livetime. The vertical lines indicate the Z/A separation between specific pairs of models of core composition. \textbf{Right:} Sensitivity bands as a function of the detector livetime, for discriminating specific pairs of models. In the right plot, the filled bands represent the discrimination power between \kj and \sakamaki\!, while the hashed bands represent \feni vs \sakamaki\!. The band limits correspond to the baseline and  optimistic assumptions on the detectors capability to filter out events which are insensitive to neutrino flavor (as explained in section~\ref{subsec:detectors}).}
\label{fig:CHI2EVOL_COMBI_NEXTGEN}
\end{figure}

Figure~\ref{fig:CHI2EVOL_UPCOMINGDETS} compares more directly the performances of the benchmark detectors in discriminating outer core composition models. Despite the differences in technology, size and reconstruction performances, we find that ORCA and DUNE reach a similar precision of 0.016 (at 1$\sigma$ C.L.) on the absolute Z/A measurement after 20 years of data taking. These results are comparable with the ones previously published~\cite{bourretVLVnT} based on a full simulation of the ORCA detector. They confirm the capability of these detectors to measure the Z/A in the Earth's outer core with a 1$\sigma$ relative precision of 3 to 4 percent. HyperKamiokande performs slightly better, with a relative precision of 2.5\%, yielding a discrimination power of approximately 0.5$\sigma$ between FeNi and FeNiH models for 25 years of measurements. A higher sensitivity can be achieved by combining the data from the three detectors into one single measurement; in that case, a 1$\sigma$ C.L. discrimination power for FeNiH vs. FeNi could be reached in 50 years of concomitant data taking, as can be seen in  figure.~\ref{fig:CHI2EVOL_COMBI_NEXTGEN}.

\subsection{A next-generation detector to identify the light elements in the outer core}
\label{subsec:res_nextgen}

While the previous result can be seen as a promising proof of concept of the method, achieving the next level of sensitivity requires both beating the statistical limitation by scaling up the detectors in size, and achieving excellent  reconstruction performances (flavor, $E_\text{reco}$, $\theta_\text{reco}$), in order to better resolve the patterns in the 2D $(E_{reco}, \theta_{reco})$ event distributions. 
 
Based on the results obtained for the upcoming detectors, we have investigated  possible realizations of such a next-generation, or NextGen, detector, that would combine the best performances of current-generation instruments, taking as benchmarks a 1 GeV detection threshold and almost perfect (99\%) discrimination between  {\it track}-like and {\it cascade-}like events.

\begin{figure}
\centering
\includegraphics[width=.49\linewidth]{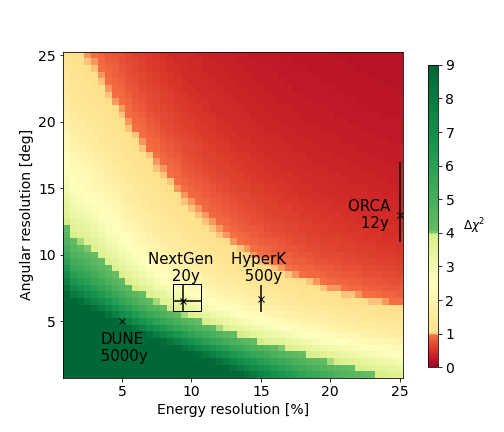}
\includegraphics[width=.49\linewidth]{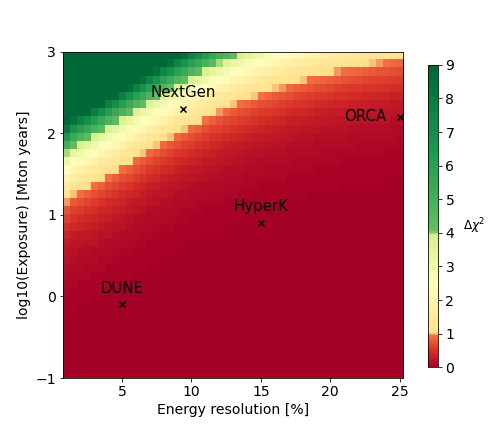}
\caption{ \textbf{$\Delta\chi^{2}$ sensitivity for discriminating \kj vs. \sakamaki with a NextGen detector for various combinations of exposure, energy and angular resolutions.} The left plot is obtained for a fixed exposure of \SI{200}~{Mton~yr}.  The color scale indicates the intervals between 1, 2 and 3\,${\sigma}$ ($68\%, 95\%, 99\% $ C.L.). Superimposed on the maps are the lifetimes required to reach the 200  Mton~yr exposure for DUNE, ORCA and HK detectors, as well as their specific resolution in the energy range of interest for the tomography measurement (3--7 GeV). The right plot is obtained by assuming a linear relationship between the energy and angular resolutions for the benchmark detectors, as suggested by the left plot.}.
 \label{fig:CHI2MAPS_ALLDET}
 \end{figure}

We performed a systematic scan of the discrimination potential of such detector over a large range of sizes, energy and angular resolutions, as presented in figure~\ref{fig:CHI2MAPS_ALLDET}. A separation power of at least $1 \sigma$ between \kj and \sakamaki models can be obtained with a rather wide combination of zenith/energy resolution parameters. A good compromise is found in the region around $\sigma(\theta)=7^\circ$ angular resolution and $\sigma(E)/E = 10\%$ energy resolution, not far from the performances associated with HyperKamiokande, but for a significantly larger, 10-Mton, detector. We provide a specific parametrization of such a NextGen detector in table~\ref{tab:detectors}. 

This scan also illustrates how the small size and limited scalability of Liquid Argon detectors is a serious obstacle to developing this technique to achieve the required exposure levels in a reasonable running time, despite their superior energy and zenith resolution. 
 Even an asymptotic evolution of the DUNE experiment, with perfect identification of all neutrino flavors and interaction channels, would only reach a $\SI{0.5}{\sigma}$ discrimination power for \kj vs. \sakamaki after 50 years running. We conclude that better perspectives for a core composition measurement arise from the water Cherenkov approach, provided that event reconstruction and identification performances similar or better to those of HyperKamiokande can be achieved in a much larger (hence likely more sparsely instrumented) detector. 
 
\begin{figure}[]
\centering
 \begin{overpic}
     [width=.49\linewidth]    {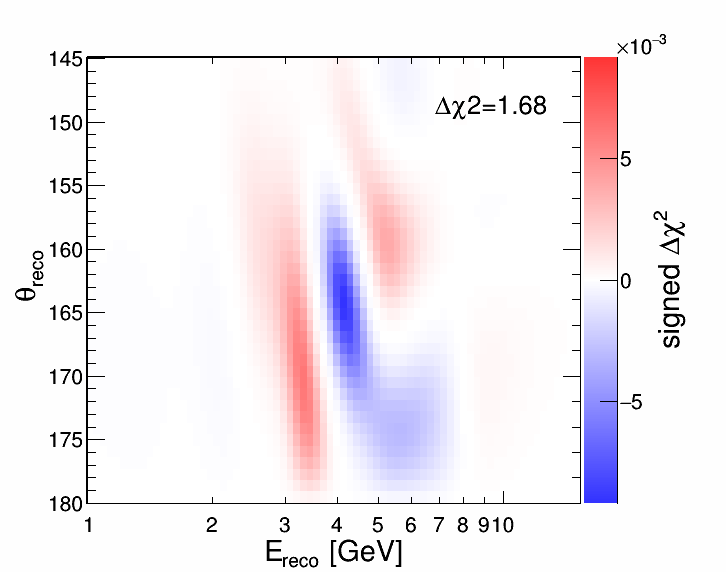}
\end{overpic}
\begin{overpic}
     [width=.49\linewidth]    {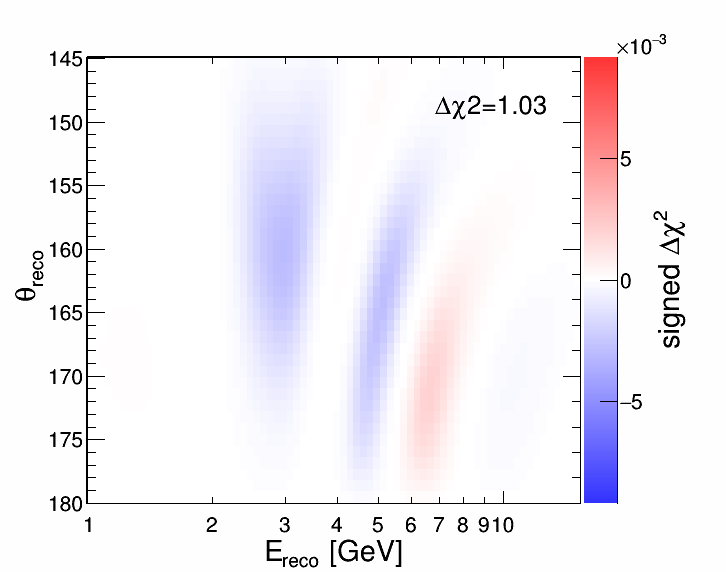}
\end{overpic}
 \caption{{\bf Sensitivity of the NextGen detector for discriminating between \kj and \sakamaki outer core compositions after 20 years data taking}. The plots show the signed $\Delta\chi^2$ maps (as defined in equation (\ref{eq:CHI2})) as a function of the reconstructed energy and zenith of the neutrinos, respectively for {\it track}-like \textbf{(left)} and  {\it cascade}-like \textbf{(right)} events.}
\label{fig:SIGCHI2_NEXT}
\end{figure}

 The performance of the NextGen detector is illustrated in~figure.~\ref{fig:SIGCHI2_NEXT} in terms of the signed $\Delta\chi^2$ maps for the discrimination between a FeNiH and a \kj composition of the outer core, for the  {\it track} and {\it cascade} channels, for 20 years of data taking. In this example, the total $\Delta\chi^2$ value is 2.71, respectively 1.68 for {\it track} and 1.03 for {\it cascade} channel. 
 
  The qualitative jump in performance that could be achieved with the NextGen detector is also evident in figure~\ref{fig:CHI2EVOL_COMBI_NEXTGEN}  where its sensitivity is compared to the one obtained from combining ORCA, DUNE and Hyperkamiokande data. The characteristics of the NextGen detector yield a sub-percent precision on the Z/A measurement, sufficient to distinguish between \kj and \sakamaki compositions at $>1 \sigma$ in a running time of about 10 years, well within the typical lifetime of a neutrino experiment (a discrimination at 90\% C.L. would be achieved in about 20 years). 

\begin{figure}[ht!]
\centering
\includegraphics[width=0.8\textwidth]{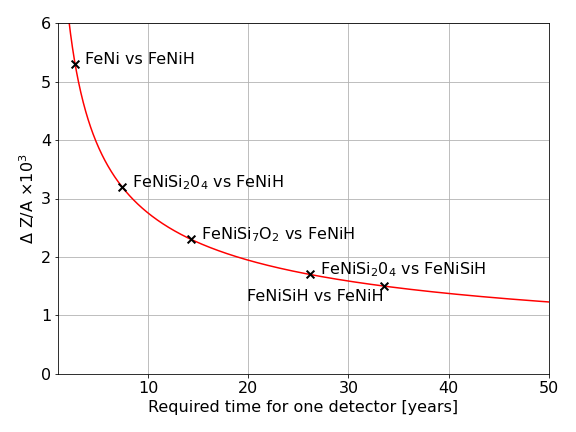}
\caption{{\bf Precision of the Z/A measurement achievable at 1$\sigma$ with the NextGen detector as a function of running time}. The crosses indicate the separation in Z/A between pairs of models considered in this study.  Not shown in the graph is the time required to distinguish FeNiSi$_7$O$_2$ vs FeNiSi$_2$O$_4$ ($\simeq 90$ years) and FeNiSi$_7$O$_2$ vs FeNiSiH ($\simeq 120$) years. If $n$ identical NextGen detectors were running in parallel, the time scale would be approximately reduced by that same factor $n$, i.e.  FeNiSi$_7$O$_2$ vs FeNiSi$_2$O$_4$ could be distinguished in about 20 years if 4 NextGen detectors would be taking data simultaneously. }
\label{fig:zoa_lim_time}
\end{figure}

We further illustrate such capabilities in figure.~\ref{fig:zoa_lim_time}, where we address the separability (at the $1 \sigma$ level) between pairs of realistic core composition models, as a function of the detector running time. Although a discrimination between composition models whose Z/A are closer than 0.001 (i.e. \badro vs \kj\!, or \kj vs. \tagawa\!) appears out of reach, our results suggest that the NextGen detector would be able to exclude or confirm the \sakamaki hypothesis against all the other realistic models considered in this study. This result can be obtained in less than 20 years if the actual outer core composition is in the family of models with an admixture of Silicium and Oxygen as only light elements.


\section{Materials and methods}
\label{sec:methods}

\subsection{Neutrino oscillations in matter}
\label{subsec:nuosc}

Neutrino flavor oscillations are a quantum-mechanical phenomenon that arises because the neutrino flavor eigenstates $\nu_e,\nu_\mu,\nu_\tau$ -- which take part in weak interactions and are therefore the observable states -- are not identical to the neutrino mass eigenstates $\nu_1, \nu_2, \nu_3$ -- which describe their propagation in vacuum. The flavor eigenstates are a quantum superposition of the mass eigenstates, whose relative phases change along the neutrino propagation path. This evolving mix of states leads to an oscillatory pattern of the detection probability of a given neutrino flavor, which depends on the neutrino energy and traveled distance.

Neutrino oscillation probabilities are calculated by solving the Schrödinger equation $i\partial_t|\nu(t)\rangle=H|\nu(t)\rangle$. The neutrino propagation Hamiltonian $H$ is approximated as the sum of two terms:

\begin{equation}
\label{eq:HAM_3NU}
    H = U\left(\begin{array}{ccc}
        0 & 0 & 0\\
        0 & \frac{\Delta{m^2_{21}}}{2E} & 0\\
        0 & 0 & \frac{\Delta{m^2_{31}}}{2E}
    \end{array}\right)U^\dagger
    +\left(\begin{array}{ccc}
        V_e & 0 & 0\\
        0 & 0 & 0\\
        0 & 0 & 0
    \end{array}\right).
\end{equation}

The first term represents intrinsic energy levels of the system in vacuum. Because neutrinos are generally observed in an ultra-relativistic context, the energy difference between mass states is approximately proportional to their difference in mass-squared $\Delta m_{ij}^2 = (m_i^2 - m_j^2)$. The Hamiltonian is shown in the flavor basis characterized by eigenstates of the weak interaction. The mixing matrix $U$ represents the unitary transformation between the mass and flavor bases and is usually parametrized in terms of three mixing angles ($\theta_{12}$, $\theta_{13}$, $\theta_{23}$) and a complex phase $\delta$, which are fundamental constants of physics (see e.g.~\cite{giganti2018} for a general discussion of oscillation formulae).

The second term in equation (\ref{eq:HAM_3NU}) arises from coherent interactions of neutrinos with electrons in the medium in which they propagate~\cite{MikheyevSmirnov,Wolfenstein}. The effective potential $V_e=\pm\sqrt{2}G_F n_e$ induces a change in the electron flavor energy level that is directly proportional to the electron number density $n_e$ and the Fermi constant $G_F$, which characterizes the strength of the weak interaction. The sign of the matter potential is positive for neutrinos and negative for antineutrinos.

The solution of the Schrödinger equation for such a system in matter of constant density amounts to computing the eigensystem of the full Hamiltonian. While in vacuum the eigenvalues and eigenvectors are given by $\Delta{m^2_{ij}}/2E$ and the matrix $U$ directly, the matter potential leads to effective eigenvalues and eigenvectors that depend on the electron density. This process can be interpreted as a modification of the fundamental parameters $\Delta{m^2_{ij}}$ and $\theta_{ij}$ induced by the neutrino environment.

For example, the effective mixing angle $\tilde{\theta}_{13}$ and the effective mass-squared splitting $\Delta\tilde{m}_{31}^2$ in matter are related to the corresponding parameters in vacuum via:
\begin{align}
  \label{eq:3NU_EFFECTIVE_PARAMS}
    \Delta\tilde{m}^2_{31} \approx \xi \cdot \Delta m^2_{31},~~~
    \sin^22\tilde{\theta}_{13} \approx \frac{\sin^22\theta_{13}}{\xi^2},
\end{align}
with a mapping parameter
\begin{align}
  \label{eq:XI}
  \xi &= \sqrt{\sin^2 2 \theta_{13} + \left(\cos 2 \theta_{13} - \frac{2\,E\,V_e}{\Delta m_{31}^2} \right)^2}.
\end{align}

At the energy and distance scales relevant to atmospheric neutrino oscillations, the probability of observing a transition between $\nu_e$ and $\nu_\mu$ flavors is proportional to $\sin^22\theta_{13}$, which is small in vacuum, as observed in oscillation measurements of reactor antineutrinos~\cite{Abe_2012,An_2012,Ahn_2012} that are insensitive to the matter potential.
However, the $\nu_e \leftrightarrow \nu_\mu$ transition probability may become large or even maximal when $\xi^2 \rightarrow \sin^2{2\theta_{13}}$, which translates into a resonance condition for the neutrino energy:
 \begin{equation}
 \label{eq:eres}
     E \rightarrow \pm \frac{\Delta m_{31}^2\cos{2\theta_{13}}}{2\sqrt{2}\,G_F\,n_e}.
 \end{equation}
 
Neutrinos traversing the Earth will experience resonant oscillations in the outer core for energies $\sim$\SI{3} GeV, and in the mantle for $\sim$\SI{7} GeV, with the exact resonant energy depending on the electron density of the material. 

\subsection{Computation of atmospheric neutrino propagation through the Earth and determination of the interacting event rate}
\label{subsec:intevents}

An almost isotropic neutrino flux is constantly produced by the interaction of cosmic rays with the atmosphere, mainly coming from the decays of charged mesons ($\pi^\pm$'s and $K^\pm$'s) and muons.  For the present study, we use as input the flux calculation produced by~\cite{Honda} for the Gran Sasso site (without mountain over the detector), averaged over all azimuth angles and assuming minimum solar activity. The flux is dominated by muon- and electron-(anti)neutrino components, with an approximate ratio of 2:1 between $\nu_\mu$ and $\nu_e$ flavors, and we have neglected the small contribution of $\nu_\tau/\bar{\nu}_\tau$. 

For a given zenith angle of incidence $\theta_z$, as defined in figure~\ref{fig:OscExample}a, the neutrino trajectory across the Earth is modelled along the corresponding baseline through a sequence of steps of constant electron density which are inferred from a radial model of the Earth with 42 concentric shells of constant $n_e$, where mass density values are fixed and follow  PREM. These shells are grouped into three petrological layers (inner core, outer core, and mantle+crust), whose composition, hence Z/A factor, is assumed to be uniform and provided in table~\ref{tab:Earth_model}). In each shell, the electron density is determined from the mass density and Z/A according to equation (\ref{eq:Z/A}). 
The probabilities of neutrino flavor transitions along their path through the Earth are computed using the \textit{OscProb} \footnote{J. Coelho {\it et al.}, \url{https://github.com/joaoabcoelho/OscProb}} package. The values of the parameters that enter the oscillation probability computation are taken from the global fit of neutrino data performed with the NuFIT analysis Version 5.0~\cite{NuFitv5.0}. We have assumed normal ordering of the neutrino mass states, {\it i.e.} $m_3 > m_1$, which is currently favoured by global fits of neutrino data.

Neutrinos with energies in the range $\sim 1$-\SI{100} GeV interact with matter mostly via scattering off nuclei, by exchanging either a neutral ($Z^0$) or a charged ($W^{\pm}$) weak-force boson that triggers a hadronic cascade. In neutral-current (NC) interactions, the neutrino survives in the scattering products and escapes. In charged-current (CC) interactions, the neutrino gives rise to a charged lepton counterpart of the same flavor ($e$, $\mu$, or $\tau$). While electrons immediately induce a secondary short electromagnetic cascade, muons travel relatively long distances, depending on their energy, before they are stopped. Because of the high mass of the $\tau$ lepton ($\approx$ \SI{1.7}{\GeV}), the contribution of (anti-)$\nu_{\tau}$-induced CC events is small in the range of energies relevant for tomography studies. For completeness, this flavor is nevertheless included in our computations. 

The rate of neutrinos interacting in a given volume of target matter is then computed, for each interaction channel (NC/CC; $e,\mu,\tau$;$\nu/\bar{\nu}$) according to equation~\ref{eq:rate}, using neutrino-nucleon cross-sections weighted for water molecules, obtained with the \texttt{GENIE} Monte Carlo neutrino generator~\cite{GENIE}.  We have neglected here the small difference in cross-section for interactions on Argon nuclei in the case of the DUNE detector.

\subsection{Detector modelling}
\label{subsec:detectors}

A neutrino interaction occurring within the detection volume will generate secondary signals that constitute the event recorded by the detector. The ability to detect, reconstruct an identify the different types of neutrino events in the energy range of interest for Earth tomography is essentially driven by the  detector size, technology and density of sensors. 

The performance of water-Cherenkov detectors is typically a trade-off between the target volume for neutrino interactions and the density of photosensors that sample the Cherenkov signal emitted by charged byproducts of the neutrino interaction. The HyperKamiokande experiment~\cite{HyperK1, HyperK2} will consist in two $\sim 200$-kton water tanks overlooked by photosensors covering its internal walls, providing  a sub-GeV threshold for neutrino detection and excellent discrimination between electron-like and muon-like signatures. To access even larger target volumes, the ORCA~\cite{KM3NeTLoI,ORCA} and PINGU~\cite{pingu} experiments propose to instrument several Mtons of seawater or polar ice with a much sparser 3D array of photosensors. 
The gain in statistics comes at the expense of a higher detection threshold (few GeV) and worse energy and direction reconstruction capabilities, which also limit their performances for neutrino flavor identification. Water-Cherenkov detectors also cannot distinguish in first approximation between neutrino and anti-neutrino events. 
    
Liquid Argon Time Projection Chambers, on the other hand, are able to reconstruct highly detailed 3D images of the neutrino event, providing excellent flavor identification capabilities, even at low (sub-GeV) energies, and energy and angular resolutions typically superior to water-Cherenkov detectors. The size of these detectors is however limited by their much higher cost. The first large-scale LArTPC detector was the ICARUS T600 detector~\cite{Rubbia:2011ft}, with an active mass of 476 tons. A new generation of LArTPCs is being developed for the DUNE experiment~\cite{DUNE-TDR}, that will include 4 detectors of 10 kton active mass each.

To simulate the detector response, we use a set of parametrized analytical functions modeling the main performance features relevant for the detection, reconstruction and classification of neutrino events:

\begin{itemize}
    \item The {\bf effective mass $M_\mathrm{eff}(E)$} is the product of the instrumented target mass $M$ of the detector and its detection efficiency, i.e. the probability for a neutrino interaction to be successfully detected as an event. $M_\mathrm{eff}$ typically increases with the neutrino energy, until reaching a plateau that saturates the fiducial mass of the detector. We have conservatively neglected here a potential increase in $M_\mathrm{eff}$ for $\nu_\mu$ CC events at high  energies, corresponding to through-going muons created in neutrino interactions outside of the detector target volume.
    The threshold for detection is mainly driven by the density (and intrinsic efficiency) of sensors. We approximate $M_{eff}(E)$ by a sigmoid function of $\log(E)$ with two adjustable parameters $E_{th}$ and $E_{pl}$, which correspond to energies where the detection efficiency reaches 5\% and 95\% respectively.
    \item The {\bf energy resolution} parameterizes the relative error on the reconstructed neutrino energy in the form of a Gaussian probability distribution function (p.d.f.) with energy-dependent width: $\sigma_E(E)/E = A_E + B_E/\sqrt{E}$.
    \item The {\bf angular resolution} parameterizes the error on the measured zenith angle in the form of a von~Mises-Fisher p.d.f. on a sphere marginalized with respect to azimuth. For wC detectors we take into account the dependence of $\sigma_{\theta}$ on the energy as $\sigma_{\theta}(E)= A_\theta + B_\theta \sqrt{E}$.
    \item The {\bf classification efficiency $\varepsilon_{class}(E)$} describes the probability for a neutrino event of energy $E$ to be correctly classified into one of the topological channels observable by the detector. We model it with a sigmoid as a function of $\log(E)$ with adjustable threshold ($E^{class}_{th}$) and plateau ($E^{class}_{pl}$) energies, maximal identification probability $P^{class}_{max}$ and minimum probability of 50\% (corresponding to the case of no separation power). 
\end{itemize}

 Events are classified according to their topological features, which depend on the signature of each interaction channel (NC/CC; $e/\mu/\tau; \nu/  \bar{\nu}$). In practice, the relative sparseness of instrumentation limits the signal sampling in the detector, and thus the reconstruction and identification performances. All experiments discussed here have a basic capability to classify events into two main observational classes: {\it tracks}, when a long muon track is observed, most of the time originating from a $\nu_\mu$ CC event; and {\it cascades}, the latter comprising $\nu_e$ CC events, but also most $\nu_\tau$ CC interactions and all NC interactions. For NC interactions, a fraction of the energy of the event is carried away by an invisible outgoing neutrino, resulting in a lower reconstructed energy and degraded angular resolution. Because NC-induced events are blind to neutrino flavor, they only decrease the experiment sensitivity. Fine-grained detectors like DUNE and HyperKamiokande have some capability to separate and reject the NC-induced events from the $\nu_e$ CC events, thereby increasing the flavor purity of the cascade channel. To account for a possible related improvement in the detector sensitivity, we have considered here both the baseline case that includes the NC-induced events in the cascade sample, and the optimistic case where the NC-induced contributions are completely removed from the sample. 
 
\begin{figure}[hb!]
    \centering
     \includegraphics[width=.49\linewidth]{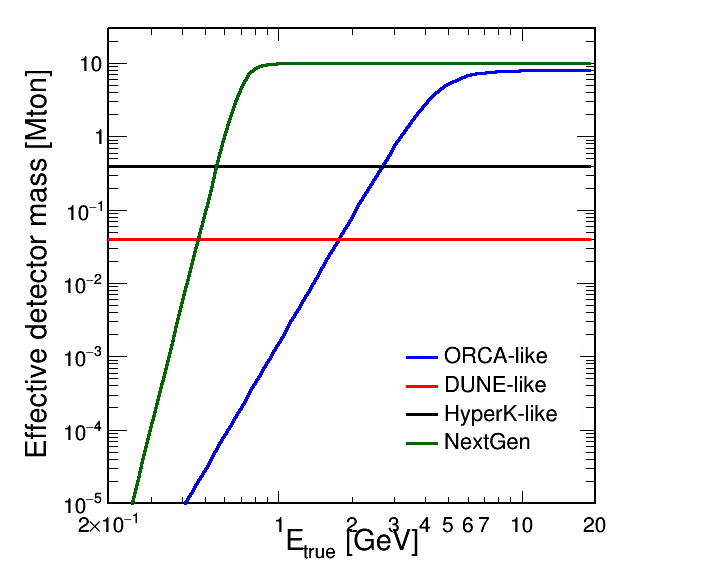}
     \includegraphics[width=.49\linewidth]{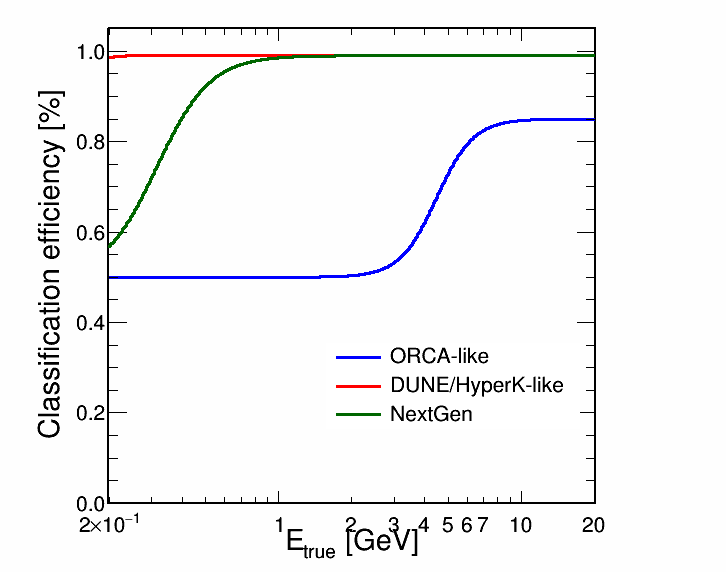}
     \includegraphics[width=.49\linewidth]{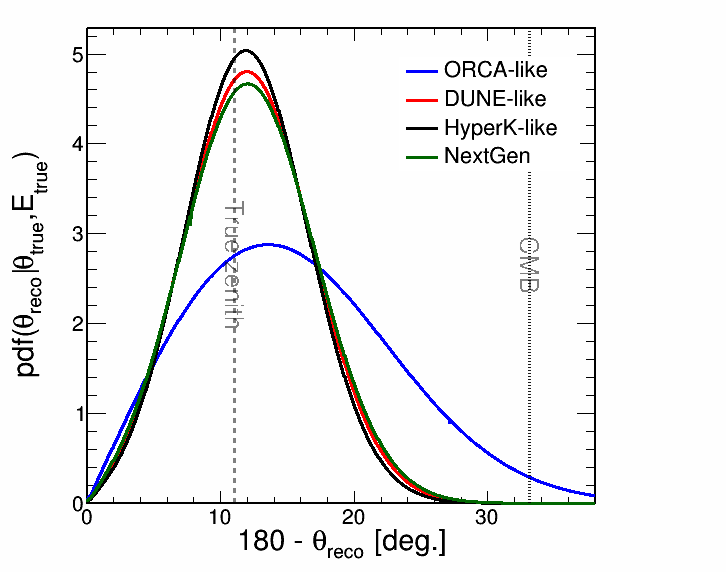}
     \includegraphics[width=.49\linewidth]{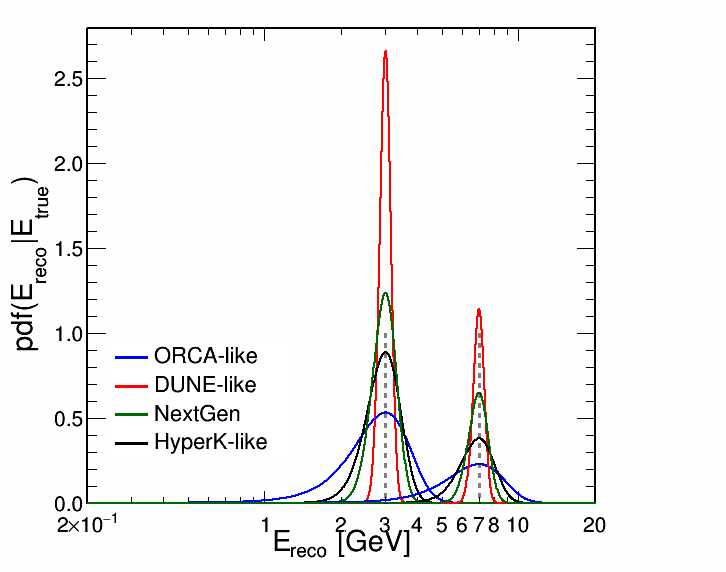}
    \caption{{\bf Examples of response functions used for the modelling of the neutrino detectors.} \textbf{Upper left:} Effective detector mass ($M_{eff}$) as a function of the neutrino true energy. \textbf{Upper right:} Classification efficiency ($\rho_{class}$) as a function of the true neutrino energy. \textbf{Lower left:} Probability distribution function for the reconstructed zenith angle for a neutrino with true energy $E_{\text{true}}=10$~GeV and true zenith angle $\theta_\text{true}\approx 169^{\circ}$ (corresponding to a neutrino trajectory grazing  the inner-outer core boundary). \textbf{Lower right:} Probability distribution function for the reconstructed energy for two specific values of the true neutrino energy ($E_{\text{true}}=1 \;\text{and}\;5$~GeV).}
    \label{fig:PARAMETERISATION}
\end{figure}

 Table~\ref{tab:detectors} provides the specific inputs chosen for different benchmark detectors representative of the upcoming generation of GeV neutrino detectors: HyperKamiokande, ORCA, and DUNE.  The DUNE-like model accounts for the superior resolution and reconstruction capabilities of the Liquid Argon detection technique, at the cost of a much smaller instrumented volume. At the other extreme, the ORCA-like model reflects the higher detection threshold and degraded reconstruction performances that result from a sparsely instrumented - although larger - volume. The HyperKamiokande-like model is a middle-way option combining a medium-size detector with a low detection threshold and good reconstruction and identification capabilities. Some examples of the corresponding response functions are presented in figure~\ref{fig:PARAMETERISATION}.   

\subsection{Computation of the expected signal}
\label{subsec:stat}

The rate of interacting neutrinos is calculated as in equation (\ref{eq:rate}); in the following we use the subscript ``true" for the quantities (energy and zenith angle) that refer to the incoming neutrino properties. This rate is then weighted with the appropriate detection efficiency at each incoming neutrino energy $E_{\text{true}}$. The expected signal in a realistic experiment is computed by a discretized convolution of the interacting rate at energy $E_{\text{true}}$ and incident at zenith angle $\theta_\text{true}$ over the E-$\cos{\theta}$ plane according to the energy and zenith resolution p.d.f.s. The reconstructed events for each interaction channel are  distributed into the two observational channels ({\it tracks} and {\it cascades}) according to the classification efficiency function $\varepsilon_\text{class}({E})$. Every bin in the final,  $(E_{\text{reco}},\theta_\text{reco})$ event oscillogram therefore (i) contains a certain fraction of misreconstructed events coming from other $(E_{\text{true}},\theta_\text{true})$ bins (ii) misses some events that end up  misreconstructed into different $(E_{\text{reco}},\theta_\text{reco})$ bins.

The final event rate expected in a given channel, in a given bin of reconstructed energy and zenith angle, is thus obtained as follows for the {\it track} channel:
\begin{eqnarray}
    R^\text{obs}_{\text{tracks}}(E_{\text{reco}},\theta_{\text{reco}}) \!\!&=& \!\!\!\!\sum_{E_{\text{true}},\theta_{\text{true}}} \left[R^{\text{int}}_\text{tracks}(E_\text{true},\theta_\text{true})\, \varepsilon_\text{class}(E_\text{true})\ + 
    R^{\text{int}}_\text{casc}(E_\text{true},\theta_\text{true})\,  \left(1-\varepsilon_\text{class}(E_\text{true})\right)\right] \nonumber \\
    & & \hspace{1.6cm}  \times PDF_\text{angle}(\theta_{\text{reco}};E_\text{true},\theta_\text{true}) \nonumber \times PDF_\text{energy}(E_{\text{reco}}; E_\text{true}) \\
    & & \hspace{1.6cm} \times \Delta E_\text{true} \times \Delta \theta_\text{true}  \times M_\mathrm{eff}(E_{\text{true}}) \label{eq:smearing} 
 \end{eqnarray}
where $PDF_\text{angle}$ and $PDF_\text{energy}$ represent the probability distribution functions of the angular and energy resolutions as defined above. A similar expression can be straightforwardly derived for the event rate in the {\it cascade} channel. 

Independently of any classical measurement error, the intrinsically probabilistic nature of neutrino interactions as a quantum process induces a statistical uncertainty on the number of events observed in each bin of $(E_{\text{reco}}, \theta_\text{reco})$. The probability to detect $N$ neutrino events of a given topology, e.g. {\it tracks}, in a predetermined interval of reconstructed energy and zenith angle is distributed according to a Poisson law, parametrized by the expected number of events $N_\text{exp} = R^\text{obs}_{\text{tracks}}(E_{\text{reco}},\theta_\text{reco}) \times T$, where $T$ corresponds to the duration of data taking. The observed number $N$ will randomly deviate from this expectation with a relative standard deviation $\sigma(N)/N_\text{exp} =1/\sqrt{N_\text{exp}}$. This uncertainty on the final measurement can only be mitigated by increasing the size of the event sample, which is proportional to the detector exposure. 

Once these fluctuations are taken into account, we  compute the log-likelihood ratio $\lambda_{LLR}$ of the resulting event histograms, which yields a measure of the significance to reject model B (the "test hypothesis") if model A (the "null hypothesis") is true. For large enough statistics, the Poisson- distributed events converge towards a normal distribution, so that $-2\lambda_{LLR}$ can be approximated by: 

\begin{equation}
    -2\lambda_{LLR} \approx \Delta\chi^{2}_i =  \frac{(n_A^i-n_B^i)^2}{n_A^i} 
\label{eq:CHI2}
\end{equation}

The total $\Delta\chi^2$ (integrated over all bins of energy and zenith angle) is related to the significance $\sigma = \sqrt{\Delta\chi^2}$, that gives the confidence level (C.L.) by which the test hypothesis can be excluded under the assumption of the null hypothesis. For illustration purposes, we present the maps in terms of the signed $\Delta\chi^2$, ie. by mutiplying the value of $\Delta\chi^2_i$ in each bin by $sign(n_A^i-n_B^i)$.


\section{Discussion}
\label{sec:discussion}

This paper presents a detailed study and comparison of the potential of upcoming atmospheric neutrino detectors to constrain the outer core composition to precision levels of relevance for geophysics. By computing the propagation and oscillation of atmospheric neutrinos through the Earth with well-established tools from the neutrino physics  community, we first confirm the theoretical capability of neutrino oscillation tomography to discriminate between different realistic core composition models. We note that the largest contribution to the theoretical signal is provided by atmospheric neutrinos reaching the detector in the $\nu_\mu$ flavor, which may explain why many of the neutrino oscillation tomography studies to date have focused on this detection channel only. 

The ultimate reach of this method, however, significantly depends on how the expected signal is effectively extracted in a realistic experiment. In order to address this question in a flexible way and with a view to the future, we have developed our analysis based on analytical parametrisations of the detector response functions, similar in concept to the studies by Winter~\cite{winter2006} and Rott {\it et al.}~\cite{rott}. This approach has allowed us to investigate for the first time different classes of experiments - namely Liquid Argon TPCs and water-Cherenkov detectors - within a unified framework. Its implementation in terms of generic  detection parameters would easily allow for further comparisons with other neutrino detection techniques that could be proposed in the future.  

One notable conclusion to be drawn from our study, is that the respective contribution of the different neutrino flavours reaching the detector to the total signal (in terms of discrimination power for different core composition models) can vary significantly for different choices of the parametrized response functions, hence for different detectors, even within the same experimental approach. This observation emphasizes the importance of considering the neutrino signal in all its components - here, namely the two observational channels {\it tracks} and {\it cascades}, in order to accurately assess the expected reach of a given experiment. 

Taking advantage of the versatility of our framework, we have further investigated the desirable detection performances for a NextGen detector that - contrarily to the upcoming generation of experiments - would be optimized for neutrino tomography. We find that such performances stem from two complementary aspects: (i) the large statistics of events provided by an experiment with 10 Mton target mass and a relatively low ($\sim$ 1 GeV) energy threshold that fully covers the energy range where resonant oscillation effects are expected for core-crossing neutrinos; and (ii) the state-of-the-art event reconstruction and classification capabilities. Limitations to this measurement arise from the intrinsic, quantum  fluctuations in the particle content of the hadronic shower that results from the fragmentation of the nucleus hit by a neutrino. These effects have been extensively discussed for ORCA-like detectors~\cite{KM3NeTfluct} and shown to limit the performances as long as individual particles in the neutrino-induced cascades cannot be reconstructed. This limitation is mitigated in instrumented water tanks \`a la  HyperKamiokande, thanks to the high density of sensors that allows a sufficient sampling of the particles produced by the neutrino interaction. 

In view of the above considerations, our favours go to the water-Cherenkov technique to define the path towards a  Next-Generation detector able to reliably test the FeNiH hypothesis for the earth's outer core composition. Such a detector should combine a multi-Megaton target volume with a sufficient density of sensors to achieve an efficient sampling  of the neutrino event topology. Because of the limited scalability of man-made water tanks, we rather advocate for a hyperdense network of light sensors deployed in natural reservoirs of water or ice. The detector topology could be either three-dimensional (à la ORCA) or two-dimensional (à la Hyperkamiokande) as long as it provides good angular coverage for neutrinos crossing the Earth's outer core. In this latter case, one possible configuration could consist in an immersed ``carpet" of densely packed photosensors looking downwards, monitoring a large target volume of water or ice. The resulting loss in sensitivity to near-horizontal neutrinos (i.e. $\theta \sim 90^\circ$) should have minimal impact on the tomography measurement: all neutrinos crossing the core will reach the detector at an angle larger than $57^\circ$ (corresponding to $\theta_Z \gtrsim 147^\circ$, and also a significant fraction of those traversing only the mantle will be detected and exploitable as a reference sample. Existing proposals for neutrino detectors made in different contexts, such as the hyper-dense 3D network Super-ORCA~\cite{SuperORCA} or the multi-megaton shallow-water network of steel tanks TITAND proposed for proton decay studies~\cite{TITAND} might also be worth reevaluating in terms of their capabilities for neutrino tomography. 

Finally, we emphasize that the timescale required for the measurements described here is inversely proportional to the total detection volume available. Deploying  multiple detector units at different positions on the surface of the globe - an effort currently being achieved with sparse wC detectors focusing on high-energy neutrino astronomy - has the extra appeal to allow for a fully 3-dimensional analysis with potentially higher scientific reach. Such a network of detectors would not only be of great interest for earth tomography purposes, but would also provide an unprecedented tool for high-statistics and high-precision studies of atmospheric neutrinos. 


\section*{Acknowledgments}

This research has benefited from the financial support of the CNRS Interdisciplinary Program 80PRIME, the LabEx UnivEarthS at Universit\'e de Paris (ANR-10-LABX-0023 and ANR-18-IDEX-0001) and the Institut Universitaire de France. 
 
 \bibliographystyle{unsrt_modjp}
\bibliography{sample}


\newpage
\section*{Tables}
\vspace*{1cm}

\begin{table}[hb!]
\centering
\begin{tabular}{lccccc}
\hline
 Label & FeNi & FeNiSi$_2$O$_4$ & FeNiSi$_7$O$_2$ & FeNiSiH  & FeNiH \\
\hline
Composition & 95 wt\% Fe & 94 wt\% Fe & 91 wt\% Fe & 93.2 wt\% Fe & 94 wt\% Fe  \\
 & 5 wt\% Ni& 5 wt\% Ni & 5 wt\% Ni & 5 wt\% Ni & 5 wt\% Ni \\
 & -  & 2 wt\% Si & 7 wt\% Si  & 6.5 wt\% Si & 1 wt\% H   \\
 & - & 4 wt\% O & 2 wt\% O & 0.3 wt\% H &  - \\
\hline
Z/A & 0.4661  & 0.4682 & 0.4691 & 0.4699 & 0.4714 \\
\hline
\end{tabular}
\caption{\textbf{Models of outer core compositions considered in this study, showing the weight fraction of the different elements and corresponding average $Z/A$.} The FeNi model is the benchmark alloy used in the present study to model the inner core composition, whereas other models introduce different combinations of light elements: FeNiSi$_2$O$_4$ \cite{badro}, FeNiSi$_7$O$_2$ \cite{KJ}, FeNiSiH \cite{tagawa} and FeNiH \cite{sakamaki2016}.  In all models the Ni content is set to 5 wt\% and Fe is the complement to 100\% once light elements have been taken into account. All elements in the table have a $Z/A$ ranging between 0.46 and 0.50, except Hydrogen whose $Z/A = 1$ is responsible for pulling the average $Z/A$ to higher values.}
\label{tab:Z/A}
\end{table}
\vspace*{1cm}

\begin{table}[hb!]
    \centering
    \begin{tabular}{lrlr}
         Layer & Shells & [$R_-,R_+$] & Z/A  \\ \hline
         \\
         Inner core & 7 & 0 - 1221.5 & 0.466 \\
         Outer core & 12 & 1221.5 - 3480.0 & 0.466 \\
         Mantle + crust & 23 & 3480.0 - 6368.0 & 0.496
    \end{tabular}
    \caption{\textbf{Compositional layers in the benchmark Earth model used in the analysis when assuming a pure FeNi outer core}. The columns indicate the number of constant density shells (each of them taking its value from PREM), the exact innermost and outermost radius (in km) and the assumed Z/A value.}
\label{tab:Earth_model}
\end{table}

\vspace*{1cm}

\begin{table}[hb!]
\hspace*{-1cm}
\footnotesize{
\begin{tabular}{lcccccccc}
 Detector & $M$ (Mton) & $E_{th}$ (GeV) & $E_{pl}$ (GeV) & $\sigma(E)/E$ & $\sigma_\theta$ (deg) & $E_{th}^{class}$ (GeV) & $E_{pl}^{class}$ (GeV) & $P_{max}^{class}$ \\
 \hline
ORCA-like& 8& 2 & 10 & 25\% & $30/\sqrt{E}$ & 2 & 10 & 85\% \\
HyperKamiokande-like & 0.40 & 0.1 & 0.2 & 15\% & $15/\sqrt{E}$ & 0.1 & 0.2 & 99\% \\
DUNE-like& 0.04 & 0.1 & 0.2 & 5\% & 5 & 0.1 & 0.2 & 99\% \\
Next-Generation & 10 & 0.5 & 1.0 & $5\%+10\%/\sqrt{E}$ & $2 + 10/\sqrt{E}$ & 0.5 & 1 & 99\% \\
\hline
\end{tabular}
}
\caption{{\bf Inputs for the response functions of the detectors considered in this study:} total target mass; threshold and plateau energy for the detection efficiency curve; energy and zenith resolutions; threshold and plateau energy for the classification efficiency curve; maximal classification probability achievable.}
\label{tab:detectors}
\end{table}


\newpage
\section*{Figures}

\end{document}